\documentclass[lettersize,journal]{IEEEtran}
\usepackage{amsmath,amsfonts}
\usepackage{algorithmic}
\usepackage{algorithm}
\usepackage{array}
\usepackage[caption=false,font=normalsize,labelfont=sf,textfont=sf]{subfig}
\usepackage{textcomp}
\usepackage{stfloats}
\usepackage{url}
\usepackage{verbatim}
\usepackage{graphicx}
\usepackage{cite}
\usepackage{color}
\usepackage[justification=centering]{caption}
\usepackage{caption}
\begin{document}

\title{Applying SDN to Mobile Networks: A New Perspective
for 6G Architecture}

\author{\IEEEauthorblockN{Rashmi Yadav\IEEEauthorrefmark{1},
Rashmi Kamran\IEEEauthorrefmark{2}, Pranav Jha\IEEEauthorrefmark{2},
Abhay Karandikar\IEEEauthorrefmark{2}\IEEEauthorrefmark{3}}\\
\IEEEauthorblockA{Department of Electrical Engineering,
Indian Institute of Technology Kanpur, India\IEEEauthorrefmark{1}, \\
\IEEEauthorblockA{Department of Electrical Engineering, Indian Institute of Technology Bombay, India\IEEEauthorrefmark{2},\\
\IEEEauthorblockA{Secretary to the Government of India,
Department of Science \& Technology, New Delhi, India\IEEEauthorrefmark{3}\\
Email: rashmiy@iitk.ac.in\IEEEauthorrefmark{1},
rashmi.kamran@iitb.ac.in\IEEEauthorrefmark{2},
pranavjha@ee.iitb.ac.in\IEEEauthorrefmark{2},
karandi@ee.iitb.ac.in\IEEEauthorrefmark{2}}}}}

\maketitle
\begin{abstract}
The upcoming Sixth Generation (6G) mobile communications system envisions supporting a variety of usage scenarios with differing characteristics, e.g., immersive communication, hyper reliable and low-latency communication, ultra massive connectivity, ubiquitous connectivity, haptic communications etc. To accommodate such diverse scenarios, the 6G system (6GS) architecture needs to be scalable, modular, and flexible. In this article, we identify some limitations of the Third Generation Partnership Project (3GPP) defined Fifth Generation System (5GS) architecture, especially that of its control plane. Further, we propose a novel architecture for the 6GS employing Software Defined Networking (SDN) technology to address these limitations. Among the different functionalities of the 5GS control plane, two key functionalities are the signalling exchange with end user devices (e.g., for user registration and user authentication) and control of user plane functions. We propose to move the “signalling handling functionality” out of the mobile network control plane and treat it as user service and signalling messages as payload or data. This proposal results in an evolved service-driven architecture for mobile networks where almost all communication with an end user (and device), including the signalling exchange, is treated as service. We explicate that the proposed architecture brings increased modularity, scalability, and flexibility to its control plane. 
We also compare its performance with the 5GS using a process algebra-based simulation tool.
\end{abstract}


\section{Introduction}
\IEEEPARstart{T}{he} notable rise in the diversity of use cases has paved the way for continued evolution of mobile networks. The upcoming 6th Generation (6G) Mobile Communication System is envisioned to support new use cases such as holographic-type communications, tactile internet, intelligent operation networks, digital twin, and Industrial Internet of Things (IIoTs) with cloudification \cite{M71}. It is also foreseen that there will be a large number of connected users in the 6G era enabled by usage scenarios like `Ubiquitous Connectivity' and `Massive Communication' \cite{M73}. 
A scalable, flexible and modular network architecture is one of the essential ingredients towards tackling the diverse usage scenarios and the anticipated massive connectivity in 6G networks. These architectural characteristics would be particularly important for the network control plane which would bear the brunt of the enormous signalling load generated by the huge number of users \cite{M78}.
\par Third Generation Partnership Project (3GPP) adopted technologies such as Network Function Virtualization (NFV), Control and User Plane Separation (CUPS), and Network Slicing (NS) for the Fifth Generation System (5GS), which resulted in improved scalability and flexibility of 5GS over the previous generation mobile communications systems. 
However, there is scope for further improvement in mobile network architecture, especially that of its control plane through the application of Software Defined Networking (SDN) technology. A survey of the existing research related to SDN-based enhancements in the mobile network is presented next. The work in \cite{M62} proposes a centralised control plane for multi-Radio Access Technology (multi-RAT) Radio Access Network (RAN) to enhance the simplicity and flexibility of the network. Relocation of the control plane functionality of RAN to the Core Network (CN) to reduce the signalling cost between RAN and core has been discussed in \cite{M61}. Authors in \cite{M64} proposed a decentralized control plane architecture for the 5GS with independent control functions for different control events for flexible and scalable networks. An SDN architecture where a middle cell and a middle cell controller are introduced between the macro cell and the small cell to reduce the control overhead of the macro cell and to address the scalability problems is proposed in \cite{M65}. In \cite{M66}, authors proposed a new 5GS core architecture based on the SDN concept. They introduced a centralised SDN controller for easier and more flexible management of the user plane. In \cite{M67}, a hierarchical control plane is designed to lighten the load of the controller. It focuses on the vertical scalability of the control plane. In \cite{M68}, a scalability metric for the SDN control plane is proposed. Besides, a comparison between different SDN architectures is analysed via mathematical methods. In \cite{M78}, authors propose to process a subset of signalling messages within the user plane (data plane). 

To summarize, current research in the context of the application of SDN technology to mobile networks mainly focuses on the centralized or distributed architecture of the control plane for reduced control overheads or scalability purposes. To the best of our knowledge, there is a limited discussion/rethink on certain other aspects, such as, what functionality should constitute the mobile network control plane within an SDN-based architecture. Is the network control plane right place for `end user signalling handling' functionality under SDN paradigm? Should `Non-Access Stratum (NAS) signalling messages' be handled by CN control plane functions such as Access and Mobility Management Function (AMF) or should this functionality be moved out of AMF? Should the user authentication function (Authentication Server Function (AUSF) in 5GS) be part of the CN control plane? These questions assume even more importance in the upcoming 6G era when increased end-user signalling load due to a surge in the number of connections may over-burden the network control plane. It should be noted that the CUPS architecture of the 5GS likely borrows its inspiration from the SDN technology \cite{M12}, \cite{M9}, \cite{M69}, though the 5GS architecture does not try to address the questions enumerated above.

In order to bring in additional enhancements to mobile network architecture, especially to its control plane, we propose to altogether separate end user (User Equipment (UE)) signalling handling from the control plane functions. In a significant departure from the existing cellular networks, the proposed architecture views `UE signalling' as payload, i.e., a form of data traversing through the cellular network, not much different from other types of data such as `video streaming' or `web browsing'. We analyse and evaluate the proposed architecture using Performance Evaluation Process Algebra (PEPA) \cite{M21}, a formal language used to model distributed systems. We also provide a comparative analysis of the proposed architecture and the 5GS architecture through example call flows for data session (Protocol Data Unit (PDU) session) establishment and mobility procedures for a UE. We find a significant reduction in the number of control messages exchanged in the proposed architecture along with an improvement in network scalability. The proposal is an extension of our earlier work \cite{M38}, which provided an initial exploration of these ideas.

The rest of the paper is organised as follows: Section \ref{lim} covers some of the limitations of the 3GPP 5GS architecture. Section \ref{arch} provides an overview of the proposed architecture, highlights its advantages and how it addresses some of the limitations of the 5GS. Section \ref{info} includes an information flow comparison of the 5GS and proposed architecture for PDU session establishment and user mobility procedures. Section \ref{model} describes the system model based on PEPA. Section \ref{perf} covers the performance evaluation. Section \ref{conc} provides the conclusion and information on future work.

\section{Limitations of 3GPP 5GS Architecture}
\label{lim}
In this section, we capture some of the limitations of the 3GPP 5GS architecture especially that of its control plane. Although there can be other limitations too, say, pertaining to the radio technology etc., those are not discussed here.

\subsection{Tight coupling of user plane control and UE signalling handling in network control plane}
The 5GS architecture supports the CUPS architecture with its control plane broadly responsible for two categories of functionalities: user plane control (or network resource control, e.g., setting up data paths for UEs via user plane) and UE signalling handling (e.g., NAS/RRC (Radio Resource Control) message exchange with UEs). There is a tight coupling between these two types of functionalities in the 5GS 
and certain CN and RAN control plane functions (e.g., AMF, gNodeB-Centralized Unit-Control Plane (gNB-CU-CP)) perform both. This may lead to control plane scalability issues in the 6G era due to expected high signalling load generated by massive connectivity \cite{M78}. A tight coupling also suggests that signalling and control can't evolve independently.
The coupling of these two types of functionalities in control plane also indicates that the 5GS architecture does not align well with the SDN paradigm. The signalling messages are exchanged with UEs to collect service requirements, e.g., requirements for PDU connectivity service and provide services such as authentication to UEs. Network functions providing such services (e.g. AUSF) may fit better within the service plane instead of the control plane in an SDN-aligned mobile network \cite{M3}.


\subsection{Non-uniform handling of services}
Services in the 5GS can be categorized into the following two types:
\begin{enumerate}
    \item Application-based services, e.g., media streaming service, IP multimedia subsystem(IMS) services, mission-critical service, Multicast/Broadcast Service (MBS).
    \item Other than these application-based services, the 5GS also provides services such as initial access, registration, authentication, PDU connectivity (connectivity to data networks via PDU sessions), and connected mode user mobility support. Such services can be called built-in (or internal) network services. 
\end{enumerate}

The two categories of services (application based services and built-in network services) are enabled differently in the 5GS. As Application (Service) Functions (AFs) are independent and decoupled from the CN and RAN functions of mobile networks, they access the control plane functions of the mobile CN over a standardized interface (via reference points: N5/N33) to enable service delivery through the user plane \cite{M12}. 
However, the delivery of built-in services is tightly integrated within the control plane of the 5GS RAN and CN; there is a limited possibility for a customer to influence it via interfaces like N33, e.g., one can't have a customer specific authentication scheme.

\subsection{Inconsistency in implementation of CUPS}
Even though 5GS has separate control and user plane functions, an altogether clean separation of functionalities between the two planes is missing. 
For example, a glaring anomaly is the transfer of the Short Message Service (SMS), a form of user data, to the UEs via control plane functions like AMF and gNB-CU-CP. SMSs are delivered using NAS signalling messages unlike other user data typically delivered via PDU sessions over user plane functions. A similar but contrasting example is that of Access Traffic steering, Switching, and Splitting (ATSSS) functionality.
To aid the ATSSS functionality, `Measurement Assistance Information', a type of signalling information is exchanged between the UE and the Performance Measurement Function (PMF), a sub-function within UPF. Even though `Measurement Assistance Information' is a type of signalling information, it is exchanged via a PDU session (i.e. via user plane functions solely) between the UE and the PMF. Hence, the mechanism is different from how other signalling information such as “RRC measurement reports” to support the ``user mobility procedure" is exchanged. 
To summarize, the 5GS does not use regular paths for both data as well as signalling exchange in certain scenarios bringing inconsistency to the CUPS implementation.

\subsection{Complex protocols between control plane and user plane}
The 5GS control plane architecture impacts the interface design between the control and user planes. For instance, F1 Application Protocol (F1AP) is the protocol used on the interface between the RAN control plane (gNB-CU-CP) and user plane (gNB-Distributed Unit (gNB-DU) or RAN-DU). It is primarily used to configure gNB-DU but also carries RRC (UE signalling) messages for UEs as the gNB-CU-CP also handles UE signalling. Integrating both these types of functionalities in a single protocol results in a relatively complex protocol between gNB-CU-CP and gNB-DU.

\subsection{Security threat to control plane}
3GPP specification \cite{M75} highlights the exposed AMF which is vulnerable to replay attacks of NAS signalling messages between the UE and AMF (a control plane function in CN). In a similar way, \cite{M77} presents the exposed RAN which is susceptible to replay attacks via RRC signalling messages exchanged between the UE and gNB-CU-CP, the control plane of 5G RAN, as the Uu interface also carries sensitive RRC signalling. Further, the European Union Agency for Cybersecurity (ENISA) \cite{M76}, in its report, highlights that the N2 interface between the 5GS RAN and AMF is a target for attackers since they carry sensitive signalling between the RAN and the CN. These scenarios highlight the ``access security threats" posed by UE signalling to the control plane functions of the 5G network, e.g., AMF, and gNB-CU-CP. 

\section{Proposed architecture for 6G System (6GS)}
 \label{arch}
 This section presents the proposed architecture with the goal to address the limitations of the 5GS (as discussed in Section \ref{lim}).
 Before delving into the details of the proposed architecture, let's grasp a basic understanding of the 5GS architecture as defined by 3GPP, especially its control plane. Please note that we exclusively present the network functions pertinent to the proposed work in this article. 
 \subsection{Introduction to the 5GS control plane}
\begin{figure}[ht]
	\centering
	\includegraphics[width=\columnwidth]{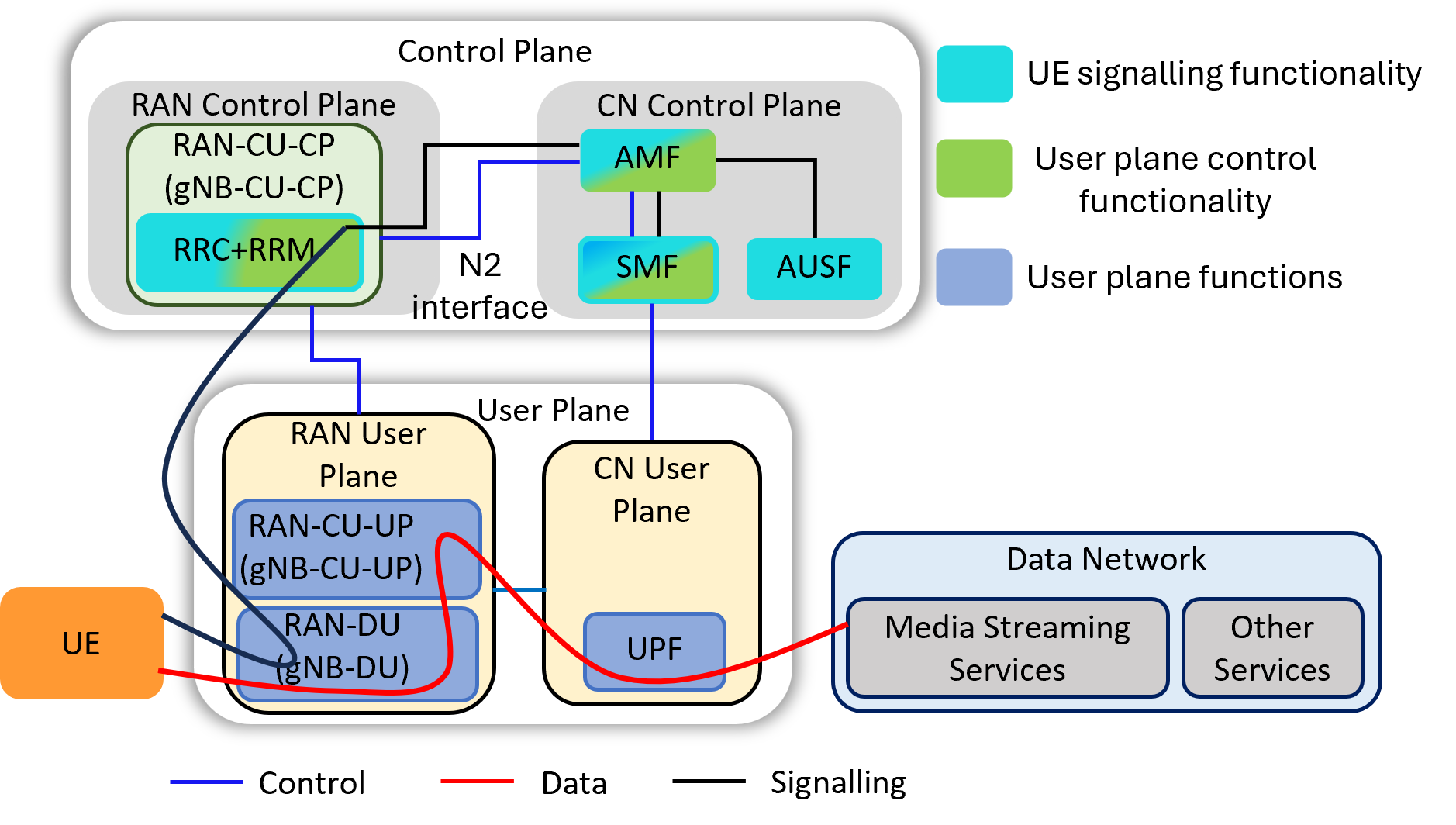}
	\vspace{-0.1cm}
    \caption{3GPP 5GS Architecture \cite{M13}.}
    \label{5GS_arch}
	\vspace{-0.5cm}
\end{figure}

 Figure \ref{5GS_arch} shows the 5GS network architecture. 5GS RAN comprises a RAN control plane (gNB-CU-CP) and a RAN user plane (gNB-DU and gNB-CU-UP) \cite{M8}.
 The control plane of the RAN, i.e., gNB-CU-CP (RAN-CU-CP) hosts the following functionality: exchange of RRC messages with UEs, RRC connection management, UE measurement reporting and control, NAS message transfer, security function, radio resource management (RRM), e.g., establishment, configuration, maintenance and release of Signalling Radio Bearers (SRBs) and Data Radio Bearers (DRBs), QoS management, connected mode mobility handling. These are broadly the RRC+RRM functionality \cite{38.300}. 

 \par Similarly, the 5GS core network consists of control plane functions, e.g., AMF, Session Management Function (SMF), AUSF and a user plane function, UPF \cite{M12}. The AMF hosts the following functionality: termination of RAN CP interface (N2), termination of NAS (N1), NAS ciphering and integrity protection, registration management,  idle mode mobility management, providing transport for Session Management (SM) messages between UE/gNB and SMF etc. The SMF includes session management, UE IP address allocation \& management, DHCPv4 and DHCPv6 functions, selection and control of UPF, configuration of traffic steering at UPF, termination of SM parts of NAS messages, etc. The AUSF supports the authentication of UEs.
\subsection{Proposed Architecture for 6GS}
 \par As illustrated in Figure \ref{5GS_arch} and previously discussed, it is evident that the control plane of both RAN (gNB-CU-CP) and CN (AMF, SMF and AUSF) broadly encompasses UE signalling handling and user plane control functionalities. This characteristic has been depicted through the usage of dual shades of colour in the figure. However, 
 we find that redefining the 5GS control plane architecture would enable a more scalable, flexible, and modular network capable of better accommodating the surge in signalling traffic in the 6G era.  

 \par
\begin{figure}[ht]
	\centering
	\includegraphics[width=1.0\columnwidth]{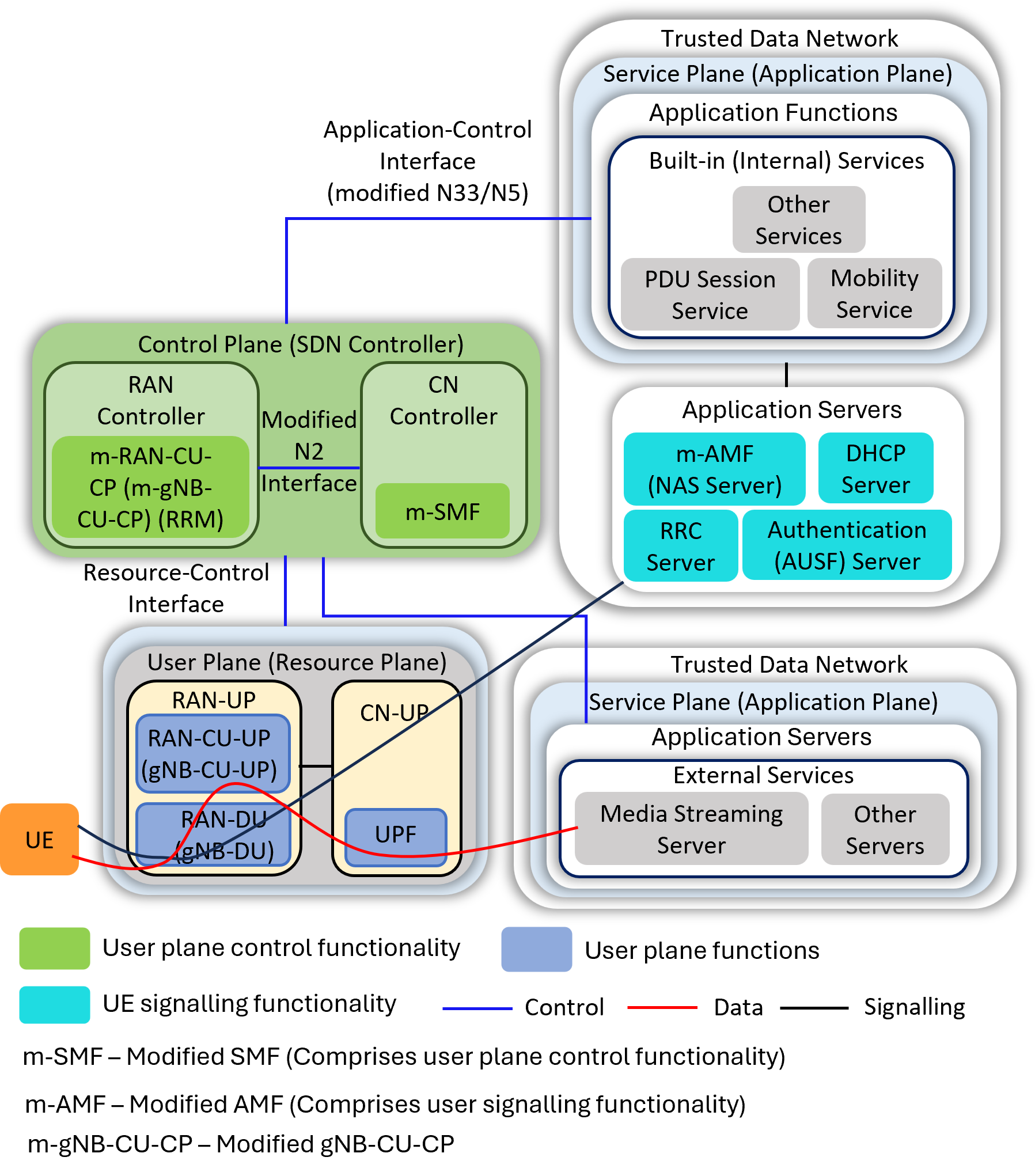}
	\vspace{-0.5cm}
    \caption{Proposed SDN-based architecture for 6G system.}
    \label{control_serv}
	\vspace{-0.2cm}
\end{figure}

 In this regard, we propose to separate the UE signalling handling from the control plane and treat it (UE signalling) as a service (data) to the user (UE). The proposal results in an evolved service-driven architecture for mobile networks where almost all communication between a user (\& UE) and the network (including signalling exchange) is treated as service, i.e., as payload or data. With the proposed separation, the control plane becomes quite thin and is left with only the user plane control functionality similar to an SDN controller, as shown in Figure \ref{control_serv}. The UE signalling handling functionality is moved out of the control plane and is now a part of the data network (service plane). Consequently, UE signalling exchange occurs between the UE and application server like functions (handling UE signalling) that are deployed as part of the data network. These application servers handling UE signalling messages are also called signalling service functions or signalling servers in this paper. 
 The communication path between a UE and the signalling servers are depicted using a solid black line in the figure. It can be called a signalling path. It is no different from regular data paths that carry user data through user plane, except for the flexibility to bypass certain user plane functions as shown in the Figure \ref{control_serv}. This split in control plane functionalities prompts a re-framing of the 3GPP 5GS architecture, leading to the redefinition of certain network functions, as detailed in the following section. 

 \par The RRC + RRM functionality, previously integrated into gNB-CU-CP (control plane of 5GS RAN), has now been split into an RRC server, a Mobility Service Function (MSF) and a RAN controller (control plane function). The following functionality is assigned to the RRC server, e.g., exchange of RRC messages with UEs, RRC connection management, security function, 
 UE measurement reporting and control, NAS message transfer. Connected mode mobility management is moved to MSF, another function in service plane.
 In contrast, user plane control functions such as establishment, configuration, maintenance and release of Signalling Radio Bearers (SRBs) and Data Radio Bearers (DRBs), QoS management, RRM etc., remain as part of the RAN controller or m-gNB-CU-CP (modified gNB-CU-CP) as shown in Figure \ref{control_serv}. Besides, the RRC server and MSF are now a part of the service plane in data network possibly alongside other service functions such as the media streaming server, as shown in Figure \ref{control_serv}. 

 \par Similarly, the control plane of the CN undergoes re-framing. The proposal involves the relocation of UE signalling handling functionalities of AMF/SMF, such as NAS signalling exchange (N1), NAS signalling security, idle mode mobility management, registration and connection management, etc. to a signalling service function, which is named as m-AMF (or NAS server) and placed in the service plane along with the RRC server. Besides, functionalities such as UE IP address allocation and management (DHCPv4, DHCPv6 functions) also is moved out of SMF and made a part of a DHCP server in service plane. PDU session management is moved to a PDU Session Service Function (PSSF) in service plane. Conversely, the remaining user plane control functionalities, including the termination of the RAN control plane interface (N2), user plane function selection and control, configuration of traffic steering at UPF, 
 etc., are retained in a modified CN controller (control plane of the core network), denoted as m-SMF. The AUSF is also moved out of the control plane to the service plane as user authentication is treated as a service in the proposed architecture.

 \par 
 Functions like PSSF or MSF are slightly different from RRC Server or AUSF as they help orchestrate the built-in services like mobility or data session establishment for UEs, which require data path (re)configuration over user plane. Like other external AF based services (e.g. media streaming) they utilize an N33/N5 like interface with the control plane to achieve the same. While functions like RRC server or AUSF directly interact with UEs via signaling paths without any interaction with the control plane. Please note that there may be separate controllers in the CN and RAN, as shown in Figure \ref{control_serv}. Further, the proposed architecture's user or resource plane may remain the same as the 3GPP 5GS with only minor changes. The proposed architecture offers many advantages discussed next.
 
 
 \subsection{Advantages of the proposed architecture}
 This section highlights a few advantages of the proposed work. Segregation of UE signalling handling functionality from the control plane \textbf{simplifies the control plane} making it thinner viz-a-viz the 5GS control plane and enhancing its simplicity, scalability and modularity.
 \par The reorganisation also aligns well with the SDN paradigm as the control plane is required to perform only user plane control functionality. The proposed architecture allows built-in services to be treated the same way as external application-based services, leading to \textbf{uniform handling of all types of services} bringing further simplicity to the design.
 \par Further, this proposal results in the simplification of the control flow. For instance, the number of sessions management-related messages is reduced due to the setup of a direct path between UE and the service functions handling UE signalling (e.g. RRC server), leading to \textbf{simplified information(call) flows}. Also, the number of hops between the RAN controller and the CN controller in the proposed architecture is less than the corresponding entities in 5GS, i.e., between gNB-CU-CP and SMF, respectively. It may result in performance improvement in terms of control plane latency. Transposition of UE signalling handling functionality to functions in service plane \textbf{simplifies the protocols} between the control pane and the user plane such as Next Generation Application Protocol (NGAP) and F1AP as these protocols are no longer required to carry UE signalling messages.
 \par The proposed architecture has a clear-cut demarcation between the user and the control plane functionality  unlike the 5GS. For example, the control plane of the proposed architecture does not participate in the transfer of user data such as SMS, as is the case with 5GS. These are handled by the Application servers (NAS/RRC servers) and the user plane functions. This reduces the complexity of the control plane further, e.g., a network function like Short Message Service Function (SMSF), used for SMS delivery in 5GS, may not be required here at all.
 \par The 5GS uses the same type of signalling messages for all use cases. However, it is possible to have different signalling requirements for different use cases, e.g., the Internet of Things (IoT) and human users. The proposed architecture may be able to support this requirement with ease by enabling deployment of \textbf{use case specific signalling servers (e.g. more than one RRC or NAS servers) to handle use case specific signalling}. In nutshell, separation of signalling handling from  the control plane enables independent evolution of signalling and control plane.  Our proposal can also support \textbf{flexible function deployment and chaining for signalling handling} as various signalling handling functions, such as the RRC server, NAS server, and Authentication server, can flexibly be placed and chained together to serve UEs. Some of these functions can also be instantiated as individual modules of a single service function depending on the requirement.
 
 \par The proposed architecture may also offer advantages in terms of network access security.  
 Since UE signalling handling is segregated from the control plane (of RAN and CN) and is terminated on a separate signalling server, it leads to the possibility of localizing the attack originating from a UE within the signalling servers without compromising the network control plane, where the logical control and management of RAN and CN are located. This aspect will be explored further in our future work.  
 \par Please note that there is no impact on the UE both with respect to signalling exchange as well as data transfer in the proposed architecture viz-a-viz the 5GS. The signalling protocol between the UE and the network may also remain the same as the 5GS, if needed. The impact is only on the network architecture and the message flow between different functions on the network side.
 

\section{Signalling and Control Information Flow Comparison}
\label{info}
In this section, we compare the signalling and control information flow of the proposed architecture and the 5GS architecture. We consider the PDU session establishment and user mobility service examples to differentiate the working of the 5GS and the proposed architectures.


\begin{figure}[ht]
	\centering
    \vspace{-0.2cm}
	\includegraphics[width=0.8\columnwidth]{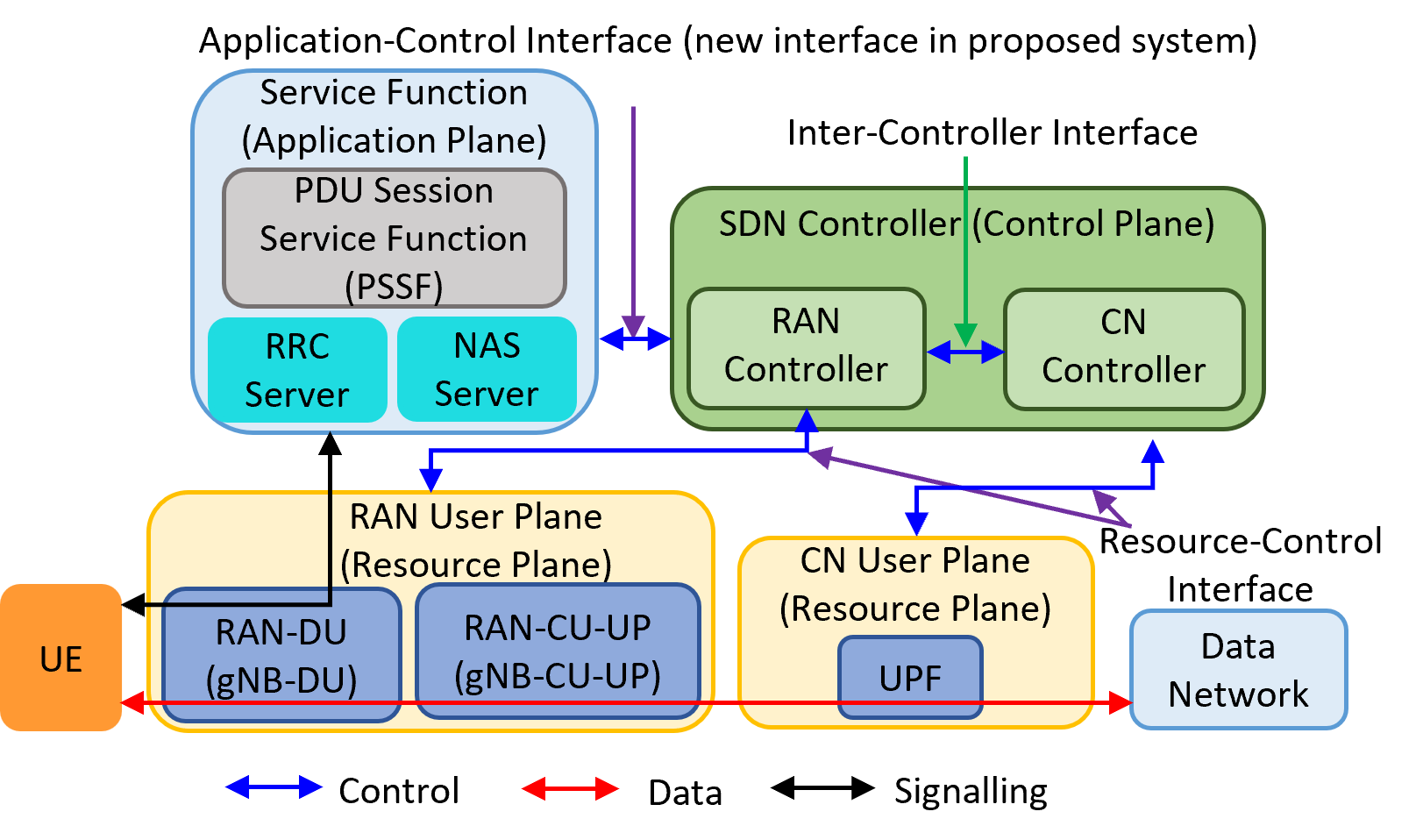}
	\vspace{-0.2cm}
    \caption{Network entities, UE signalling and control message flow for PDU session establishment in the proposed architecture.}
    \vspace{-0.1cm}
    \label{PDU_serv}
	\vspace{-0.5cm}
\end{figure}

\subsection{PDU session establishment}
\label{pdu}
Figure \ref{PDU_serv} shows the entities involved in control flow and signalling exchange for PDU session establishment for the proposed architecture. In 5GS, messages are exchanged between UE and SMF for PDU session-related signalling via RAN (it requires gNB-DU and gNB-CU-CP) and AMF. It implies that in the 5GS, signalling messages pass through multiple hops. 
However, as shown in Figure \ref{PDU_serv}, signalling message exchange between UE and the service function (PSSF) requires only one hop (only RAN-DU) in the proposed architecture. The RRC server, NAS server and the PDU Session service function can be instantiated as modules of a single network entity, reducing the number of hops required. Further, the control plane receives all UE specific requirements from the PSSF via the application-control interface (N33/N5 like) and establishes the PDU session over the user plane.

\begin{figure}[ht]
	\centering
	\vspace{0.1cm}
	\includegraphics[width=\columnwidth]{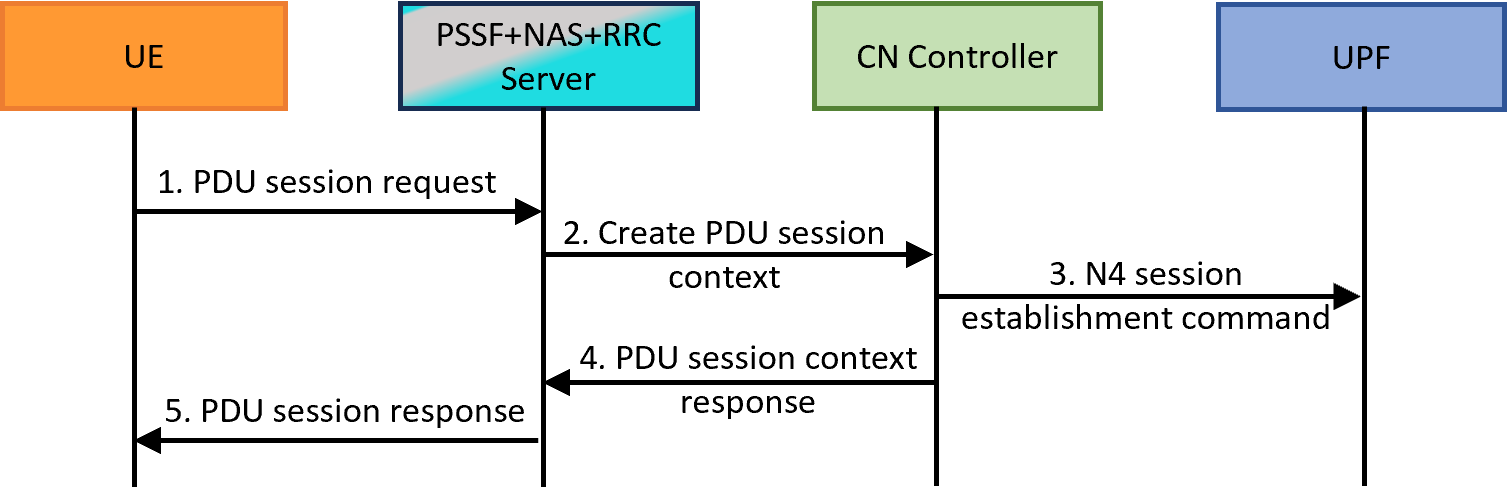}
    \vspace{-0.3cm}
	\caption{PDU session establishment procedure in the proposed architecture.} 
    \label{pdu_seq} 
	\vspace{-0.4cm}
\end{figure}

The complete message sequences for establishing PDU sessions for the 5GS are detailed in \cite{M13} while simplified call flow for the proposed architecture is shown in Figure \ref{pdu_seq}\footnote{In call flows and simulations, only those messages are considered and compared which are different in the proposed and 5GS architectures}. Please note that the controllers do not require response messages from the user (resource) plane, as the controller is fully aware of the user plane resources; hence, such response messages have been eliminated from the proposed architecture. For example, the N4 session modification request and response are exchanged between SMF and UPF in 5GS architecture \cite{M13}, while the session establishment command (message 3 in Figure \ref{pdu_seq}) sent by the CN controller to the CN user plane (UPF) in the proposed architecture does not need a response from the UPF. Such reductions in messages simplify both the session establishment and user mobility procedures (to be discussed next). 
\textit{Please note that even though RAN-User Plane (RAN-UP) and other RAN functions/messages are also necessary, we have shown only the CN functions in the call flow to keep the analysis tractable. However, keeping the RAN functions and the associated interactions out of the call flows is not likely to alter the conclusions drawn here.}

\subsection{User mobility}
\label{mob}
\begin{figure}[ht]
	\centering
	\includegraphics[width=0.8\columnwidth]{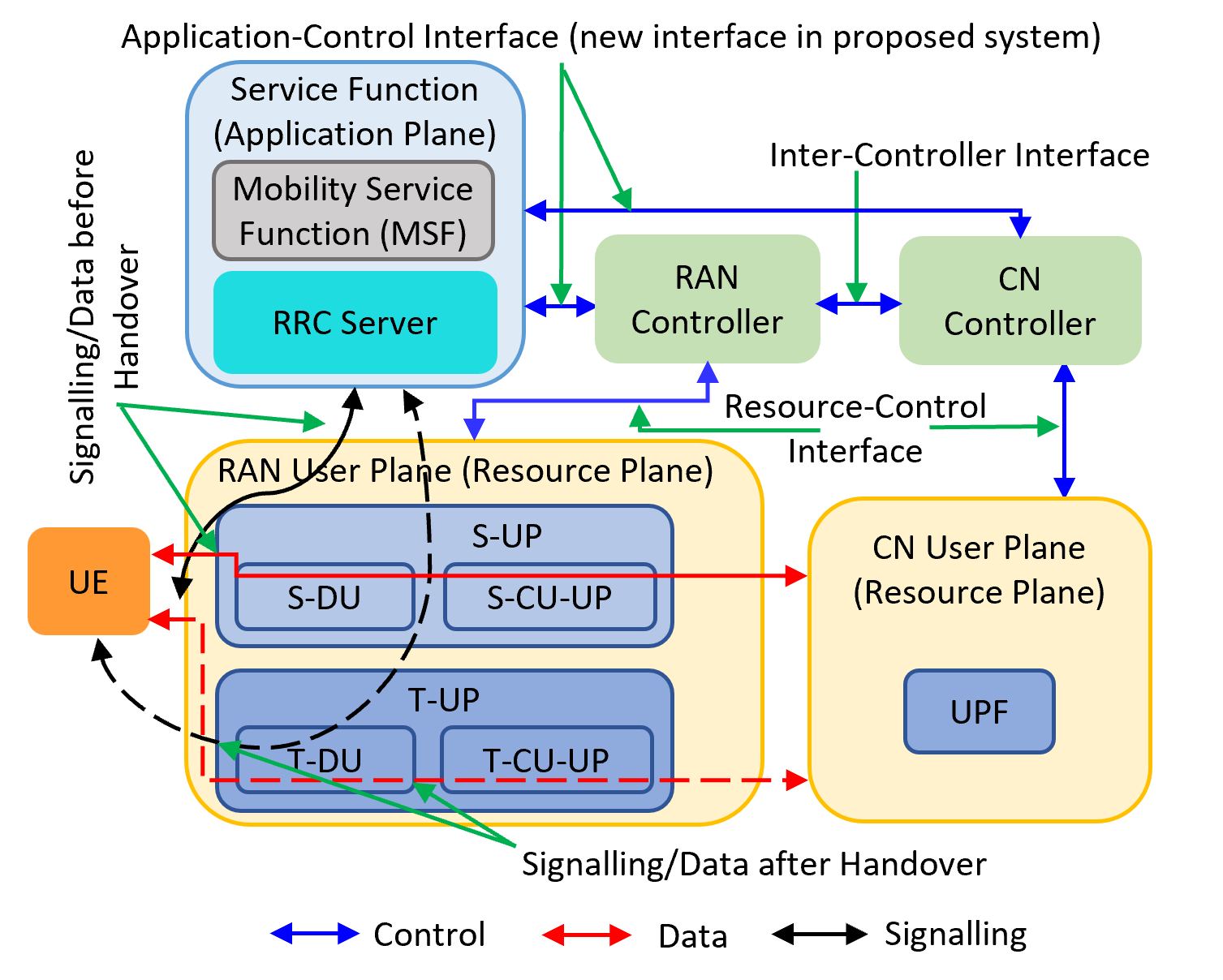}
	\vspace{-0.1cm}
    \caption{Network entities, UE signalling and control message flow in case of user mobility for the proposed architecture.}
    \vspace{-0.1cm}
    \label{mob_serv}
	\vspace{-0.1cm}
\end{figure}

We consider user mobility as another service to illustrate the difference between the 5GS and the proposed architecture in terms of control flow and signalling exchange. Figure \ref{mob_serv} shows the network entities, signalling and control message flow for the user mobility service of the proposed architecture. S-DU and T-DU represent source gNB-DU and target gNB-DU, respectively. Similarly, the Source-Centralized Unit-User Plane (S-CU-UP) and Target-Centralized Unit-User Plane (T-CU-UP) represent source gNB-CU-UP and target gNB-CU-UP, respectively. 
Without any loss of generalization, a single RAN Controller is used to control both source and target RAN user plane. Also, the interaction between the RAN controller and the CN controller is through the inter-controller interface, as shown in Figure \ref{mob_serv}.


\begin{figure}[ht]
	\centering
	\vspace{-0.1cm}
	\includegraphics[width=1.0\columnwidth]{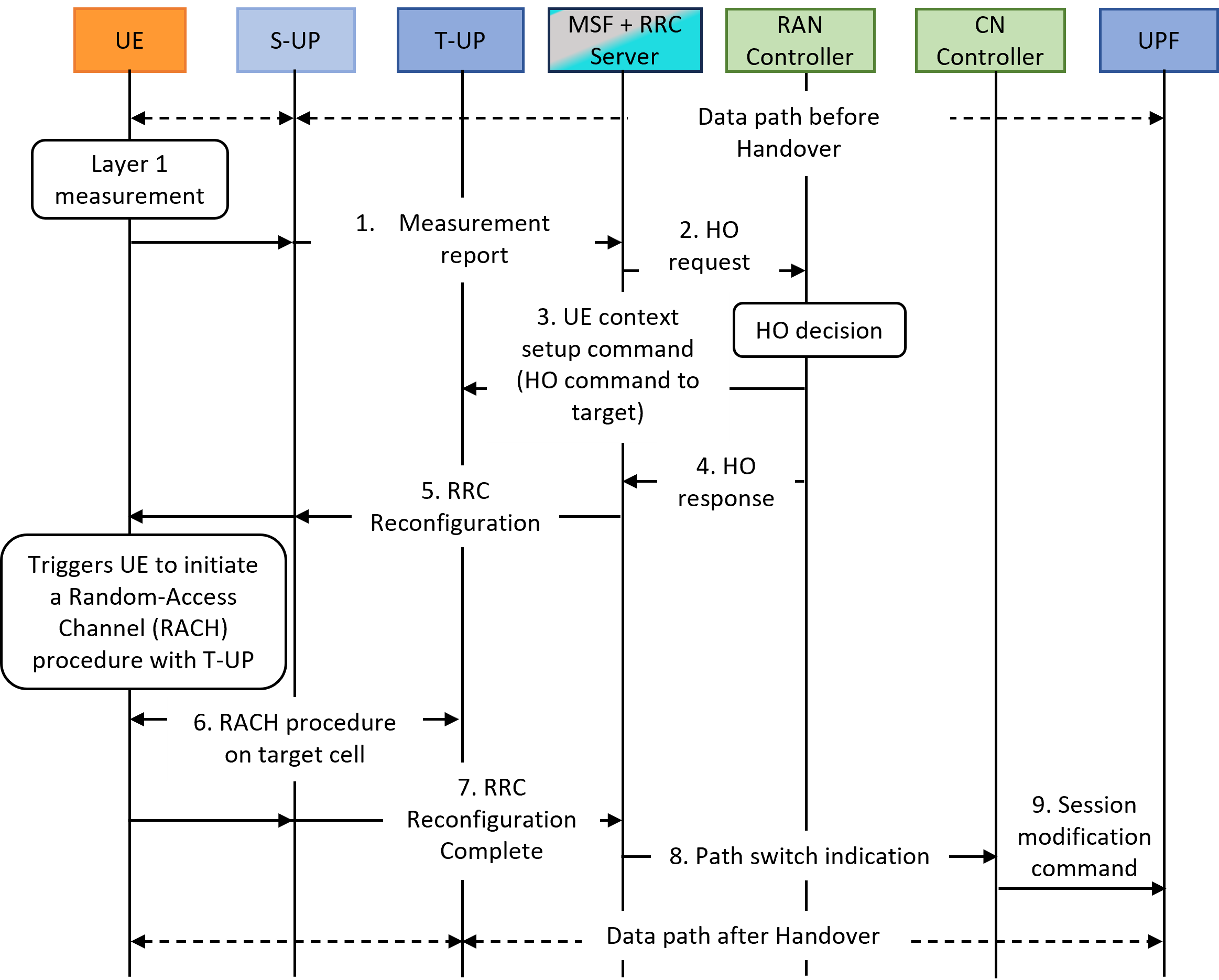}
	\vspace{-0.3cm}
	\caption{User mobility procedure in the proposed architecture.} 
    \label{infoserv}
	\label{interface2}
	\vspace{-0.5cm}
\end{figure}

User mobility call flow for the 5GS is available in \cite{M13}. Figure \ref{infoserv} here shows the user mobility call flow which illustrates the mobility procedure of the proposed architecture. For the sake of simplicity, splitting S-UP into S-DU and S-CU-UP and T-UP into T-DU and T-CU-UP is not shown in the call flow. Similar to the previous call flow, we show the MSF and the RRC server in a single box. The reasons behind the simplification of user mobility procedure are two fold. One is the same as explained for PDU session establishment in Section \ref{pdu}. The other is the simplicity brought in by the proposed architecture to support built-in services. It views mobility simply as a two step process, i.e., setup of new data paths and termination of the old paths.   


\label{PDU_flow}


\section{\textbf{System Model}}
\label{model}

\begin{table}
\caption{PDU session establishment}\label{tab:table1}
\centering
\vspace{0.1cm}
\fontsize{7.5pt}{7.5pt}\selectfont
\begin{tabular}{|p{0.19\columnwidth}|p{0.6\columnwidth}|}
\hline
\textbf{PEPA Modules} & \textbf{\centering Code Description}\\
\hline
UE NF & \textit{$Ue_1$ $_{=}^{def}$ ($acc_{uep}$, $r_a$).(process, $r_{iat}$).$Ue_2$} \\ 
 & \textit{$Ue_2$ $_{=}^{def}$ ($req_{pduse}$, $r_r$).($rep_{pduse}$, $r_r$).$Ue_1$}\\
\hline
PSSF NF & \textit{$Pssf_1$ $_{=}^{def}$ ($req_{pduse}$, $r_r$).$Pssf_2$} \\ 
           & \textit{$Pssf_2$ $_{=}^{def}$ ($acc_{pssfp}$, $r_a$).(process, $r_p$).$Pssf_3$} \\ 
           & \textit{$Pssf_3$ $_{=}^{def}$ ($req_{sc}$, $r_r$).($rep_{sc}$, $r_r$).$Pssf_4$} \\ 
           & \textit{$Pssf_4$ $_{=}^{def}$ ($acc_{pssfp}$, $r_a$).(process, $r_p$).$Pssf_5$} \\ 
           & \textit{$Pssf_5$ $_{=}^{def}$ ($rep_{pduse}$, $r_r$).$Pssf_1$} \\
\hline
CN Controller & \textit{$Con_1$ $_{=}^{def}$ ($req_{sc}$, $r_r$).$Con_2$} \\ 
NF & \textit{$Con_2$ $_{=}^{def}$ ($acc_{conp}$, $r_a$).(process, $r_p$).$Con_3$} \\ 
& \textit{$Con_3$ $_{=}^{def}$ ($req_{n4est}$, $r_r$).($rep_{n4est}$, $r_r$).$Con_4$} \\ 
& \textit{$Con_4$ $_{=}^{def}$ ($acc_{conp}$, $r_a$).(process, $r_p$).$Con_5$} \\ 
& \textit{$Con_5$ $_{=}^{def}$ ($rep_{sc}$, $r_r$).$Con_1$} \\
\hline
UPF NF & \textit{$Upf_1$ $_{=}^{def}$ ($req_{n4est}$, $r_r$).$Upf_2$} \\ 
& \textit{$Upf_2$ $_{=}^{def}$ ($acc_{upfp}$, $r_a$).(process, $r_p$).$Upf_1$} \\ 
\hline
UE Processor & \textit{$Uep_1$ $_{=}^{def}$ ($acc_{uep}$, $r_a$).$Uep_2$} \\ 
& \textit{$Uep_2$ $_{=}^{def}$ (process, $r_p$).$Uep_1$} \\
\hline
PSSF Processor & \textit{$Pssfp_2$ $_{=}^{def}$ (process, $r_p$).$Pssfp_1$} \\
\hline
CN Controller & \textit{$Conp_1$ $_{=}^{def}$ ($acc_{conp}$, $r_a$).$Conp_2$} \\ 
Processor & \textit{$Conp_2$ $_{=}^{def}$ (process, $r_p$).$Conp_1$} \\
\hline
UPF Processor & \textit{$Upfp_1$ $_{=}^{def}$ ($acc_{upfp}$, $r_a$).$Upfp_2$} \\ 
& \textit{$Upfp_2$ $_{=}^{def}$ (process, $r_p$).$Upfp_1$} \\
\hline
System & ((($Ue_1$[n] $_{L_1}^{\bowtie}$ $Pssf_1$[$N_{pssf}$.$N_{pssfp}$.$N_t$]) \\  
Equation & $_{L_2}^{\bowtie}$ $Con_1$[$N_{con}$.$N_{conp}$.$N_t$]) \\  
& $_{L_3}^{\bowtie}$ $Upf_1$[$N_{upf}$.$N_{upfp}$.$N_t$]) \\  
& $_{L_4}^{\bowtie}$ ((($Uep_1$[n] $_{\phi}^{\bowtie}$ $Pssfp_1$[$N_{pssf}$.$N_{pssfp}$])  \\ 
& $_{\phi}^{\bowtie}$ $Conp_1$[$N_{con}$.$N_{conp}$]) \\
& $_{\phi}^{\bowtie}$ $Upfp_1$[$N_{upf}$.$N_{upfp}$]) \\
\hline
Cooperation & $L_1$ = $<$$req_{pduse}$, $rep_{pduse}$$>$ \\ 
Set & $L_2$ = $<$$req_{sc}$, $rep_{sc}$$>$ \\ 
& $L_3$ = $<$$req_{n4est}$$>$ \\ 
& $L_4$ = $<$$acc_{uep}$, $process$, $acc_{pssfp}$, \\ 
& $acc_{conp}$, $acc_{upfp}$$>$ \\ 
& \textit{$\phi$} = $<>$ \\
\hline
\end{tabular}
\vspace{-0.5cm}
\end{table}

This section presents the system model for the proposed architecture using PEPA. PEPA is a formal high-level language for the quantitative modelling of a distributed system \cite{M21}. 
Table \ref{tab:table1} and Table \ref{tab:table2} represent the system model for the proposed architecture for the PDU session establishment and user mobility procedures, respectively. To explain the system models, we use the PDU session establishment (or session establishment) and user mobility procedure (as shown in Figure \ref{pdu_seq} and Figure \ref{infoserv}). 

The session establishment procedure requires PSSF, CN controller and UPF as the key CN functions in the proposed architecture. These NFs are modelled as PEPA components. In addition, a UE is also modelled as a PEPA component. Each PEPA component (representing UE or a CN NF) goes through a set of states during the handling of the procedure. The individual component states are denoted by associating a unique number with the name of the component (e.g., $Pssf_1$ represents the first state of the component, PSSF). Each component performs a set of actions, such as accessing the processor or sending a request/response. These actions are denoted in lowercase, and subscripts are added to provide further distinction (as $action_{actiondetail}$). For example, the notation for the action of PDU session establishment request and response can be $req_{pduse}$ and $rep_{pduse}$, respectively. Each action is associated with a specific rate value, $r$. The rate (number of actions performed per unit time) models the expected duration of the action in the PEPA component, and its values for different actions are taken as reference from \cite{M19}, \cite{M17} and \cite{M20}.

Let us now understand the details of modelling of NF states as shown in Table\,\ref{tab:table1}. Consider UE as an example. The UE acquires the processor ($acc_{uep}$, $r_a$) in its initial state, $Ue_1$, and performs the processing action ($process$, $r_{iat}$) before sending a request. The second state, $Ue_2$, models the request ($req_{pduse}$, $r_r$) and response ($rep_{pduse}$, $r_r$) messages exchanged between UE and PSSF for the PDU session establishment.  
NFs acquire processors to process a request/response. In Table\,\ref{tab:table1}, UEP, PSSFP, CONP and UPFP are the processing entities for UE, PSSF, CN controller (CON) and UPF respectively. These processing entities are modelled such that each NF processor has two states. For instance, the first state of UEP, $Uep_1$, is for acquiring the processor ($acc_{uep}$), and the second state, $Uep_2$, performs the processing action ($process$). Similarly, the other NFs and their processing entities are modelled.

\begin{table}
\caption{User mobility}\label{tab:table2}
\centering
\vspace{0.1cm}
\fontsize{7.5pt}{7.5pt}\selectfont
\begin{tabular}{|p{0.18\columnwidth}|p{0.7\columnwidth}|}
\hline
\textbf{PEPA Modules} & \textbf{\centering Code Description}\\
\hline
UE NF & \textit{$Ue_1$ $_{=}^{def}$ ($acc_{uep}$, $r_a$).($measure$, $r_{iat}$).$Ue_2$} \\ 
& \textit{$Ue_2$ $_{=}^{def}$ ($reconfig$, $r_r$).$Ue_3$} \\ 
& \textit{$Ue_3$ $_{=}^{def}$ ($rachreq$, $r_r$).($rachres$, $r_r$).$Ue_4$} \\ 
& \textit{$Ue_4$ $_{=}^{def}$ ($reconfigcomp$,$r_r$).$Ue_1$} \\
\hline
T-UP NF & \textit{$Upt_1$ $_{=}^{def}$ ($pathsetup$, $r_r$).$Upt_2$} \\ 
& \textit{$Upt_2$ $_{=}^{def}$ ($acc_{uptp}$, $r_a$).($process$,$r_p$).$Upt_3$} \\ 
& \textit{$Upt_3$ $_{=}^{def}$ ($rachreq$,$r_r$).($rachres$,$r_r$).$Upt_1$} \\
\hline
MSF NF & \textit{$Msf_1$ $_{=}^{def}$ ($measure$,$r_r$).$Msf_2$}\\ 
& \textit{$Msf_2$ $_{=}^{def}$ ($acc_{msfp}$,$r_a$).($horeq$,$r_r$).$Msf_3$} \\ 
& \textit{$Msf_3$ $_{=}^{def}$ ($hores$,$r_r$).$Msf_4$} \\ 
& \textit{$Msf_4$ $_{=}^{def}$ ($acc_{msfp}$,$r_a$).($reconfig$,$r_r$).$Msf_5$} \\ 
& \textit{$Msf_5$ $_{=}^{def}$ ($reconfigcomp$,$r_r$).$Msf_6$} \\ 
& \textit{$Msf_6$ $_{=}^{def}$ ($acc_{msfp}$,$r_a$).($pathswitch$,$r_r$).$Msf_1$} \\
\hline
RAN & \textit{$Ran_1$ $_{=}^{def}$ ($horeq$,$r_r$).$Ran_2$} \\ 
Controller NF & \textit{$Ran_2$ $_{=}^{def}$ ($acc_{ranp}$,$r_a$).($pathsetup$,$r_r$)} \\ 
& .($hores$,$r_r$).$Ran_1$ \\
\hline
CN & \textit{$Cn_1$ $_{=}^{def}$ ($pathswitch$,$r_r$).$Cn_2$} \\ 
Controller NF & \textit{$Cn_2$ $_{=}^{def}$ ($acc_{cnp}$,$r_a$).($session$,$r_r$).$Cn_1$} \\
\hline
UPF NF & \textit{$Upf_1$ $_{=}^{def}$ ($session$,$r_r$).$Upf_2$} \\ 
& \textit{$Upf_2$ $_{=}^{def}$ ($acc_{upfp}$,$r_a$).($process$,$r_p$).$Upf_1$} \\
\hline
UE Processor & \textit{$Uep_1$ $_{=}^{def}$ ($acc_{uep}$,$r_a$).$Uep_2$} \\ 
& \textit{$Uep_2$ $_{=}^{def}$ ($measure$,$r_{iat}$).$Uep_1$} \\
\hline
T-UP & \textit{$Uptp_1$ $_{=}^{def}$ ($acc_{uptp}$,$r_a$).$Uptp_2$} \\ 
Processor & \textit{$Uptp_2$ $_{=}^{def}$ ($rachreq$,$r_r$).$Uptp1$ +($rachres$,$r_r$).$Uptp_1$} \\
\hline
MSF & \textit{$Msfp_1$ $_{=}^{def}$ ($acc_{msfp}$,$r_a$).$Msfp_2$} \\ 
Processor & \textit{$Msfp_2$ $_{=}^{def}$ ($horeq$,$r_r$).$Msfp_1$+($reconfig$,$r_r$)} \\ 
& .$Msfp_1$+($pathswitch$,$r_r$).$Msfp_1$ \\
\hline
RAN & \textit{$Ranp_1$ $_{=}^{def}$ ($acc_{ranp}$,$r_a$).$Ranp_2$} \\ 
Processor & \textit{$Ranp_2$ $_{=}^{def}$ ($pathsetup$,$r_r$).($hores$,$r_r$).$Ranp_1$} \\
\hline
CN Processor & \textit{$Cnp_1$ $_{=}^{def}$ ($acc_{cnp}$,$r_a$).$Cnp_2$} \\ 
& \textit{$Cnp_2$ $_{=}^{def}$ ($session$,$r_r$).$Cnp_1$} \\
\hline
UPF Processor & \textit{$Upfp_1$ $_{=}^{def}$ ($acc_{upfp}$,$r_a$).$Upfp_2$} \\ 
& \textit{$Upfp_2$ $_{=}^{def}$ ($session$,$r_r$).$Upfp_1$} \\
\hline
System & ((((($Ue_1$[n]$_{L_1}^{\bowtie}$$Upt_1$[$N_{upt}$.$N_{uptp}$.$N_t$]) \\ 
Equation & $_{L_2}^{\bowtie}$$Msf1$[$N_{msf}$.$N_{msfp}$.$N_t$])\\ 
& $_{L_3}^{\bowtie}$$Ran_1$[$N_{ran}$.$N_{ranp}$.$N_t$]) \\ & $_{L_4}^{\bowtie}$$Cn_1$[$N_{cn}$.$N_{cnp}$.$N_t$]) $_{L_5}^{\bowtie}$$Upf_1$[$N_{upf}$.$N_{upfp}$.$N_t$]) \\ & $_{L_6}^{\bowtie}$((((($Uep_1$[n]$_{\phi}^{\bowtie}$$Uptp_1$[$N_{upt}$.$N_{uptp}$]) \\ & $_{\phi}^{\bowtie}$$Msfp_1$[$N_{msf}$.$N_{msfp}$]) $_{\phi}^{\bowtie}$$Ranp_1$[$N_{ran}$.$N_{ranp}$]) \\ & $_{\phi}^{\bowtie}$$Cnp_1$[$N_{cn}$.$N_{cnp}$]) $_{\phi}^{\bowtie}$$Upfp_1$[$N_{upf}$.$N_{upfp}$]) \\
\hline
Cooperation & $L_1$ = $<rachreq, rachres>$ \\
Set & $L_2$ = $<$$measure$, $reconfig$, $reconfigcomp$$>$ \\
& $L_3$ = $<pathsetup, horeq, hores>$ \\
& $L_4$ = $<pathswitch>$ \\
& $L_5$ = $<session>$ \\ 
& $L_6$ = $<$$acc_{uep}$, $acc_{uptp}$, $acc_{msfp}$, \\ 
& $acc_{ranp}$, $acc_{cnp}$, $acc_{upfp}$$>$ \\
& $\phi$ = $<>$ \\
\hline
\end{tabular}
\vspace{-0.7cm}
\end{table}

As discussed in this section, the system model uses the following additional parameters: $n$ denotes the number of UEs; $N_{pssf}$, $N_{con}$, and $N_{upf}$ are the number of NF instances for PSSF, CN controller (CON), and UPF, respectively. Similarly, $N_{pssfp}$, $N_{conp}$, and $N_{upfp}$ are the number of PSSF processors (PSSFPs), CN controller processors (CONPs) and UPF processors (UPFPs), respectively. Please note that each processor can handle a set of concurrent threads, $N_t$. Thus, the product $N_{nf}$·$N_{nfp}$·$N_t$ (where $N_{nf}$ are the number of NFs, $N_{nfp}$ are the number of processors for each NF as mentioned in the system model equation) represents the total number of threads for a type of NF. Moreover, the product $N_{nf}$·$N_{nfp}$ is the total number of processors allocated to a type of NF, e.g., for UPF processor.

The system equation represents the overall system model. The cooperation operator (``${\bowtie}$''), for example, A $_{L}^{\bowtie}$ B, models the interactions between NFs A and B over the actions defined in the cooperation set $L$. It can be noted that it is possible that component A $_{L}^{\bowtie}$ B will have different behaviour from component A $_{K}^{\bowtie}$ B if L$\neq$K. Let us consider an example from Figure \ref{pdu_seq}, where PSSF and CN controller (CON) interact with each other for `create session context request/response' $req_{sc}$/$rep_{sc}$. These actions are defined in cooperation set $L_2$, as shown in Table\,\ref{tab:table1}. Therefore, the system equation $Pssf_1$[$N_{pssf}$.$N_{pssfp}$.$N_t$] $_{L_2}^{\bowtie}$ $Con_1$[$N_{con}$.$N_{conp}$.$N_t$], models the interaction between PSSF and CN controller over the cooperation set $L_2$. In a similar way, the overall system equation, as shown in  Table\,\ref{tab:table1} and Table\,\ref{tab:table2} represents the interaction between the various NFs as shown in the two call flows,  Figure \ref{pdu_seq} and Figure \ref{infoserv}, respectively.

\section{performance evaluation}
\label{perf}
 This section presents the performance comparison between the 5GS and the proposed architecture analysed using the PEPA Eclipse plug-in \cite{M10}, a software tool integrated into the popular Eclipse platform. As mentioned before, the control plane performance has been evaluated here. 
    \newline \textbf{Session establishment rate }: The number of sessions established per unit time, measured for the action, $rep_{pduse}$, which describes the completion of the session establishment procedure. Similarly, to assess the performance of user mobility service, the \textbf{number of successful handovers} is measured for the message session modification command (message 9 in Figure \ref{infoserv}) signifying the completion of the user mobility procedure.
    \newline  \textbf{Average response time}: It measures the UE waiting time for any specific request, e.g., `session establishment' and reflects the system's operating speed. In this case, we consider the average response time as the duration required to complete the session establishment procedure. Similarly, we consider the user mobility procedure's average response time as the duration to complete the user mobility procedure.
    \newline  \textbf{Processor utilisation}: Processor utilisation measures the NFs processor capacity utilised during a procedure. The utilisation of any NF processor (for example, \textit{PSSF} processor) while performing any procedure is derived from its population level analysis (one of the features available in the tool) \cite{M79}.
    \newline \textbf{Scalability}: Scalability \textit{(S)}, in simple terms, measures a network's ability to increase or decrease its size, performance and cost in response to changes in system processing demands. Alternatively, according to Equation \ref{eq:eq1}, scalability can be defined as the ratio between the productivity of a system at two configurations  (configuration here implies the number of NFs used) having different scales, say \textit{$m_1$} and \textit{$m_2$} \cite{M16}, which corresponds to the different numbers of NFs used in the network, say \textit{$m_1$} = \textit{(1,1,1)} and \textit{$m_2$} = \textit{(3,3,1)}. \textit{$m_1$} and \textit{$m_2$} details are discussed in subsection \ref{sec:sec6}.  
    The mathematical expression for scalability is given as \cite{M16}:
    \begin{equation}\label{eq:eq1}
S(m_1,m_2) = \frac{C(m_2)}{C(m_1)},
\end{equation}  

Where, \textit{C(m)} is the productivity of a system at the scale \textit{m}, given by (Equation \ref{eq:eq2}): 
\begin{equation}\label{eq:eq2}
C(m) = \frac{t(m)\cdot r(m)}{U(m)},
\end{equation}

Where \textit{t(m)} is the average number of sessions established at scale \textit{m}, \textit{U(m)} is the processor utilisation of the system (as defined in (3) of Section \ref{perf}) at scale \textit{m}, and \textit{r(m)} (Equation \ref{eq:eq3}) is determined by evaluating the response time performance of the scaled system. We consider the following equation \cite{M16} to evaluate the performance function \textit{r(m)} by using the average response time \textit{T(m)}, at scale \textit{m}, with the target average response time \textit{T} \cite{M17}. 

\begin{equation}\label{eq:eq3}
r(m) =\frac{1}{1+T(m)/T}.
\end{equation}

\subsection{\textbf{Results and Analysis}}
In this section, we present the performance results for 5GS and the proposed architecture in the case of PDU session establishment service and user mobility service. 

\subsubsection{PDU Session Establishment Service}
\label{sec:sec6}
\begin{figure}[ht]
	\centering
	\includegraphics[width=0.8\columnwidth]{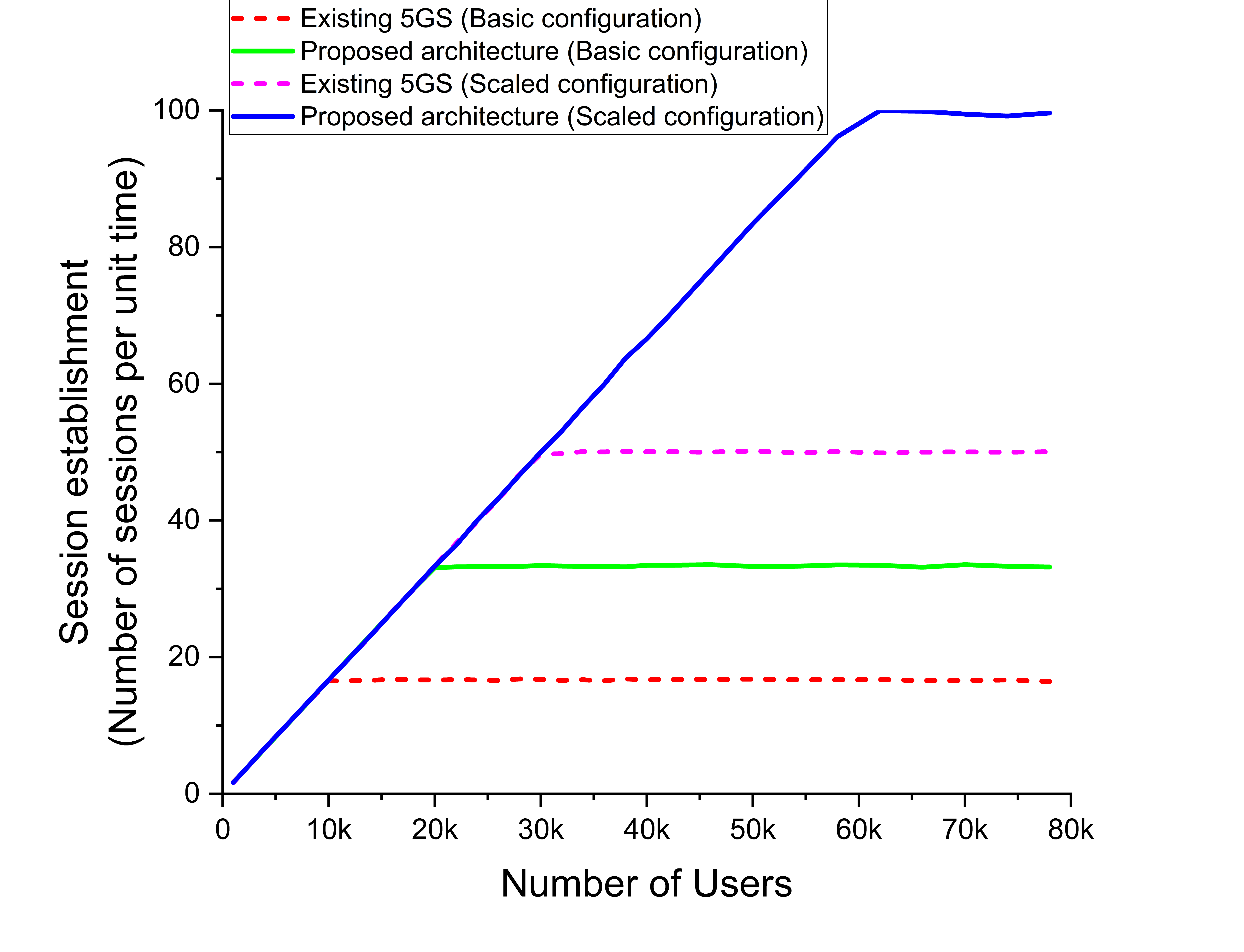}
	\vspace{-0.4cm}
    \caption{Session establishment rate for the proposed and the 5GS architecture.}
    \vspace{-0.1cm}
    \label{fig:fig5}
	\vspace{-0.2cm}
\end{figure}

The performance analysis of the proposed architecture and the 5GS for the session establishment procedure is discussed in this section. Figure \ref{fig:fig5} shows the session establishment rate with respect to the number of UEs for 5GS and the proposed architecture using two different configurations. For instance, \textit{($N_{pssf}$, $N_{con}$, $N_{upf}$)} = \textit{(1,1,1)} for the proposed architecture is the basic configuration (\textit{$m_1$}) with single NF instances assigned to each NF, i.e., to PSSF, CON, UPF and \textit{($N_{pssf}$, $N_{con}$, $N_{upf}$)} = \textit{(3,3,1)} is the configuration for a scaled system (\textit{$m_2$}) with three NF instances assigned to PSSF and CON while one to UPF. Similarly, basic and the scaled configuration for 5GS is defined as \textit{($N_{amf}$, $N_{smf}$, $N_{upf}$)} = \textit{(1,1,1)} and \textit{($N_{amf}$, $N_{smf}$, $N_{upf}$)} = \textit{(3,3,1)}, respectively.  

\begin{figure}[ht]
	\centering
	\includegraphics[width=0.8\columnwidth]{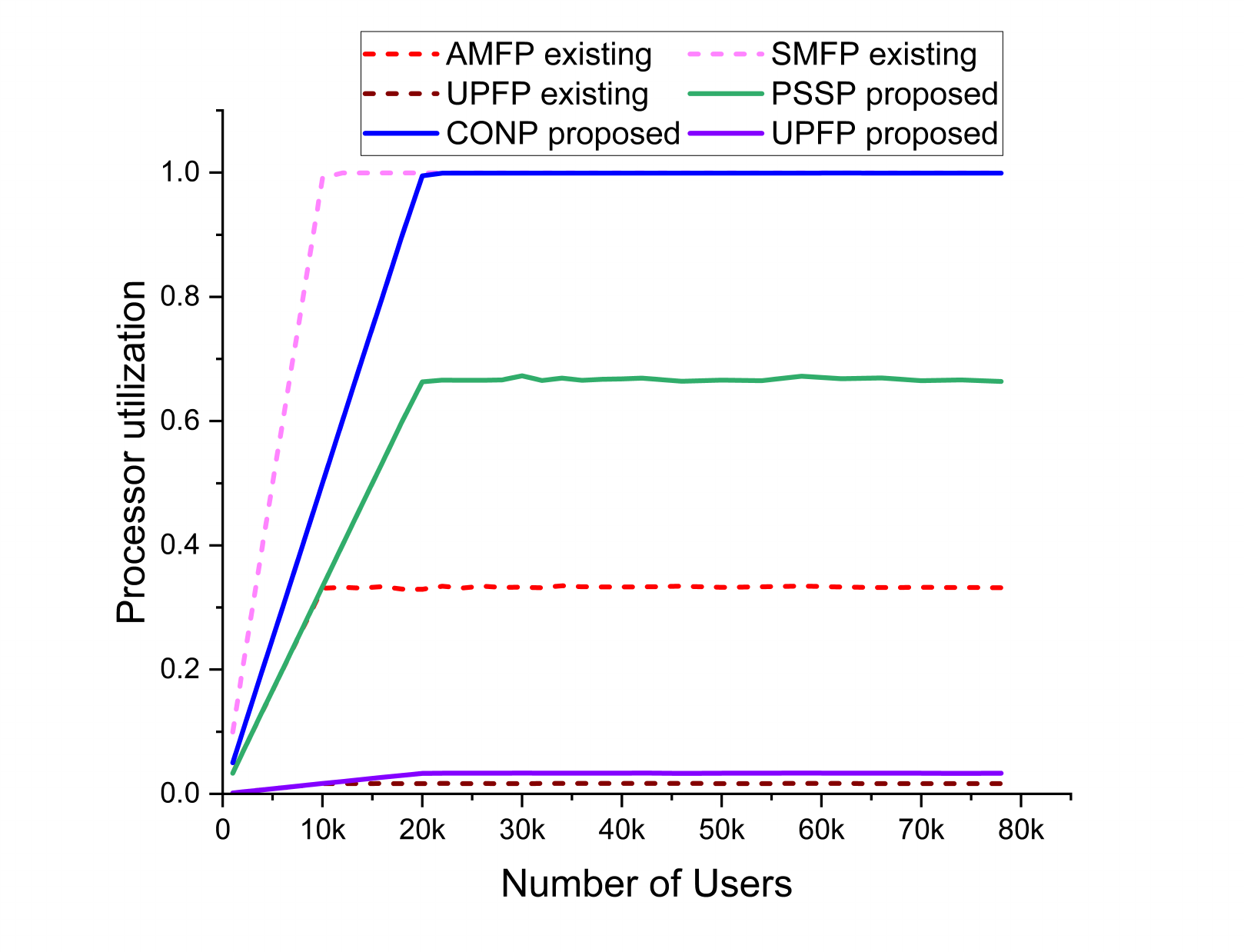}
	\vspace{-0.4cm}
    \caption{Processor utilisation of session establishment for the proposed and the 5GS architecture with the basic configuration.}
    \vspace{-0.1cm}
    \label{fig:fig9}
	\vspace{-0.1cm}
\end{figure}

\begin{figure}[ht]
	\centering
	\includegraphics[width=0.8\columnwidth]{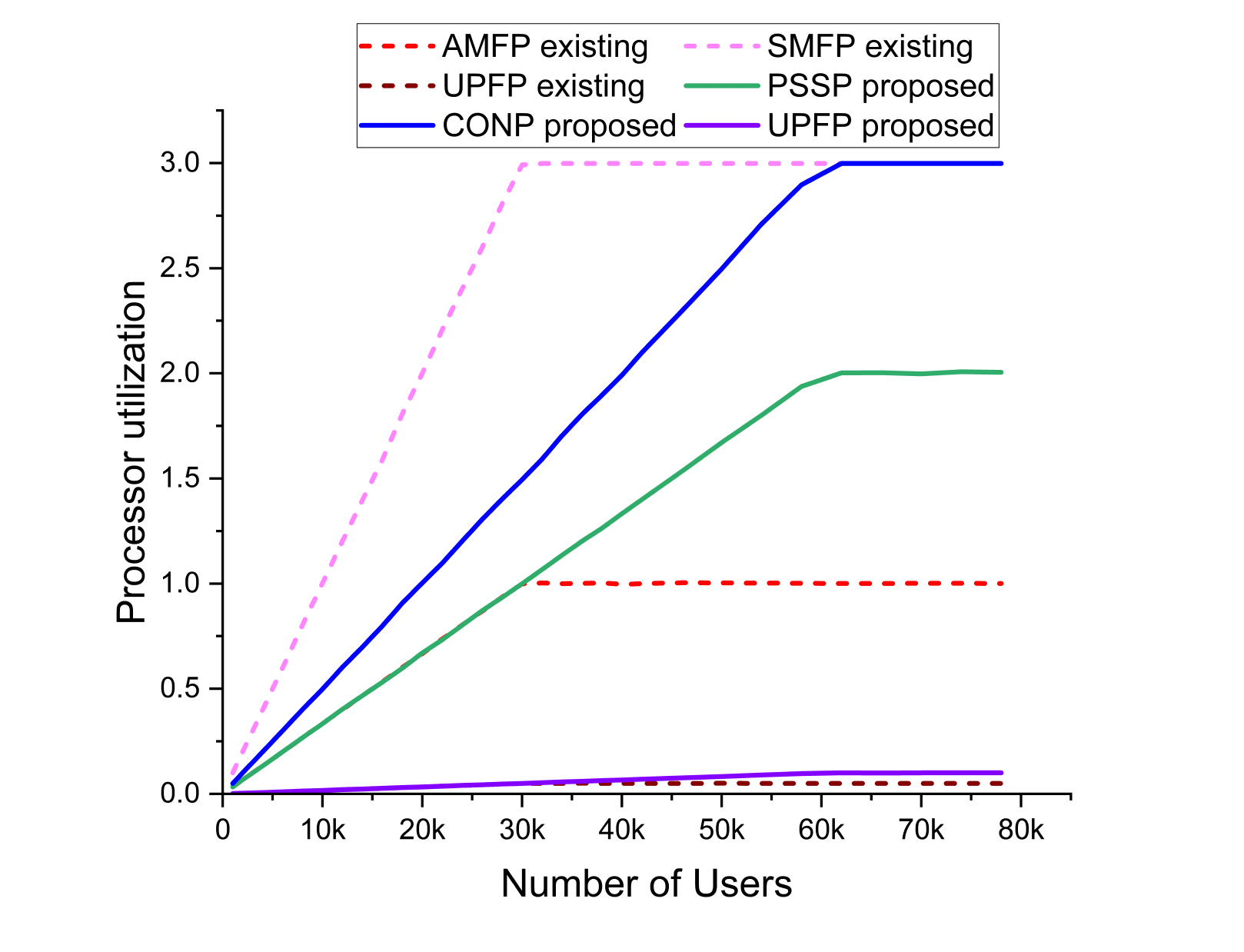}
	\vspace{-0.4cm}
    \caption{Processor utilisation of session establishment for the proposed and the 5GS architecture with scaled configuration.}
    \vspace{-0.1cm}
    \label{fig:fig10}
	\vspace{-0.3cm}
\end{figure}

Results show that the proposed architecture can achieve a higher session establishment rate compared to the 5GS in case of both basic and scaled configurations. Although the session establishment rate has increased using a scaled configuration for both the proposed and the 5GS architectures compared to the session establishment rate achieved using a basic configuration, the proposed architecture achieves a higher session establishment rate than the 5GS. 
The saturation point for 5GS, as shown in Figure \ref{fig:fig5}, is around 10,000 users i.e. it can serve a maximum number of 10,000 users in case of basic configuration, while the session establishment rate for the proposed architecture saturates at around 20,000 users. However, the 5GS saturates at around 34,000 users in scaled configuration whereas the proposed architecture saturates at 62,000 users as shown in Figure \ref{fig:fig5}. As the saturation point is reached, the network starts dropping the incoming requests from users. This means that with the given number of processors/NFs, the proposed architecture can achieve a higher session establishment rate. 
The processor utilisation for all NFs of the 5GS and the proposed architecture for basic and the scaled configuration are shown in Figure \ref{fig:fig9} and Figure \ref{fig:fig10}, respectively. It should be observed that the saturation point for processor utilisation is much higher for the proposed architecture viz-a-viz the 5GS. For instance, the PSSFP reaches its maximum utilisation explaining the saturation point for the session establishment rate. However, at this point, CONP and UPFP are not fully utilised. These results show that the request processing chain fails if an NF becomes a bottleneck for the consecutive chain. 

\begin{figure}[ht]
	\centering
	\includegraphics[width=0.8\columnwidth]{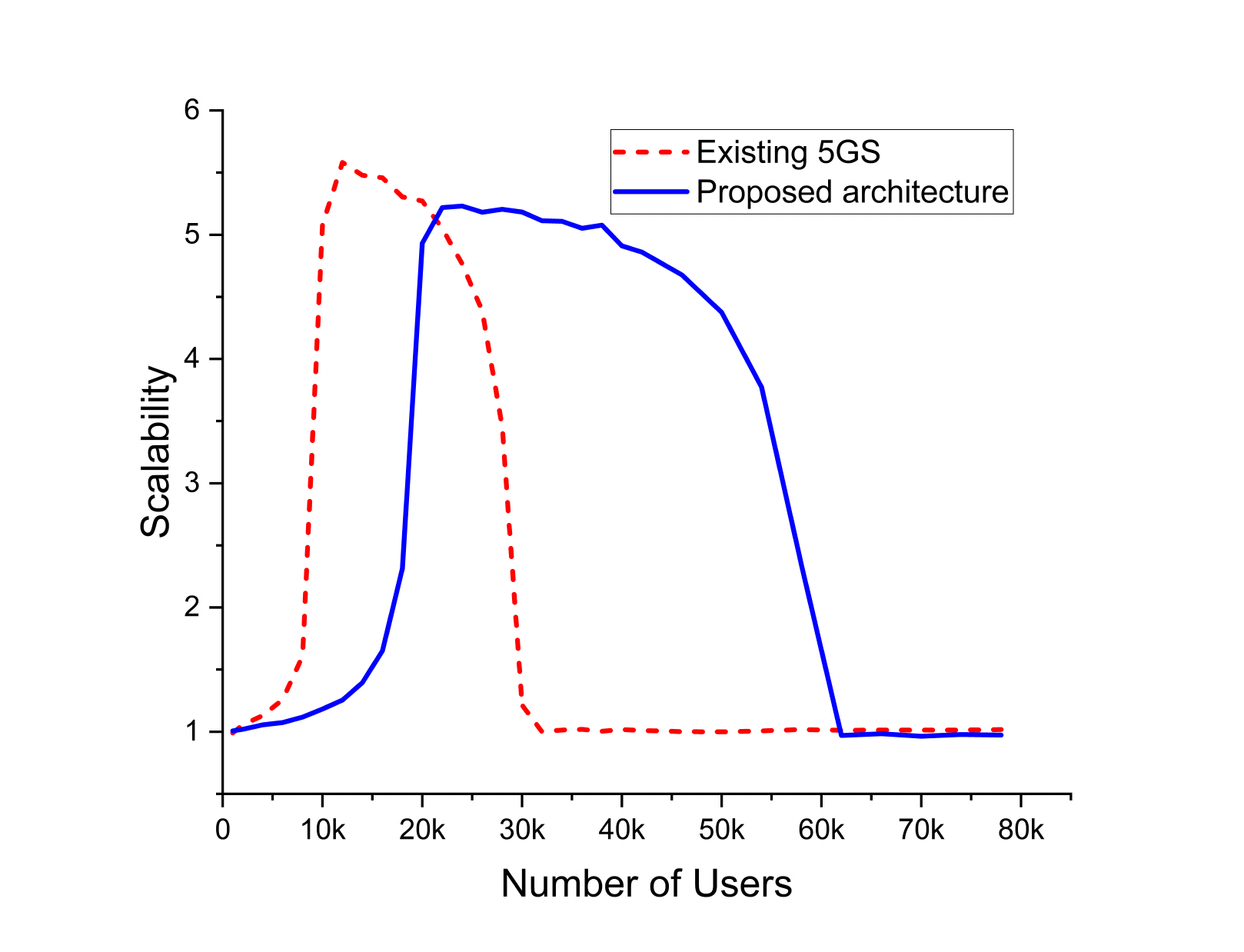}
	\vspace{-0.4cm}
    \caption{Scalability in case of PDU session establishment for the proposed and the 5GS architecture.}
    \label{fig:fig11}
	\vspace{-0.4cm}
\end{figure}

\begin{figure}[ht]
	\centering
	\includegraphics[width=0.8\columnwidth]{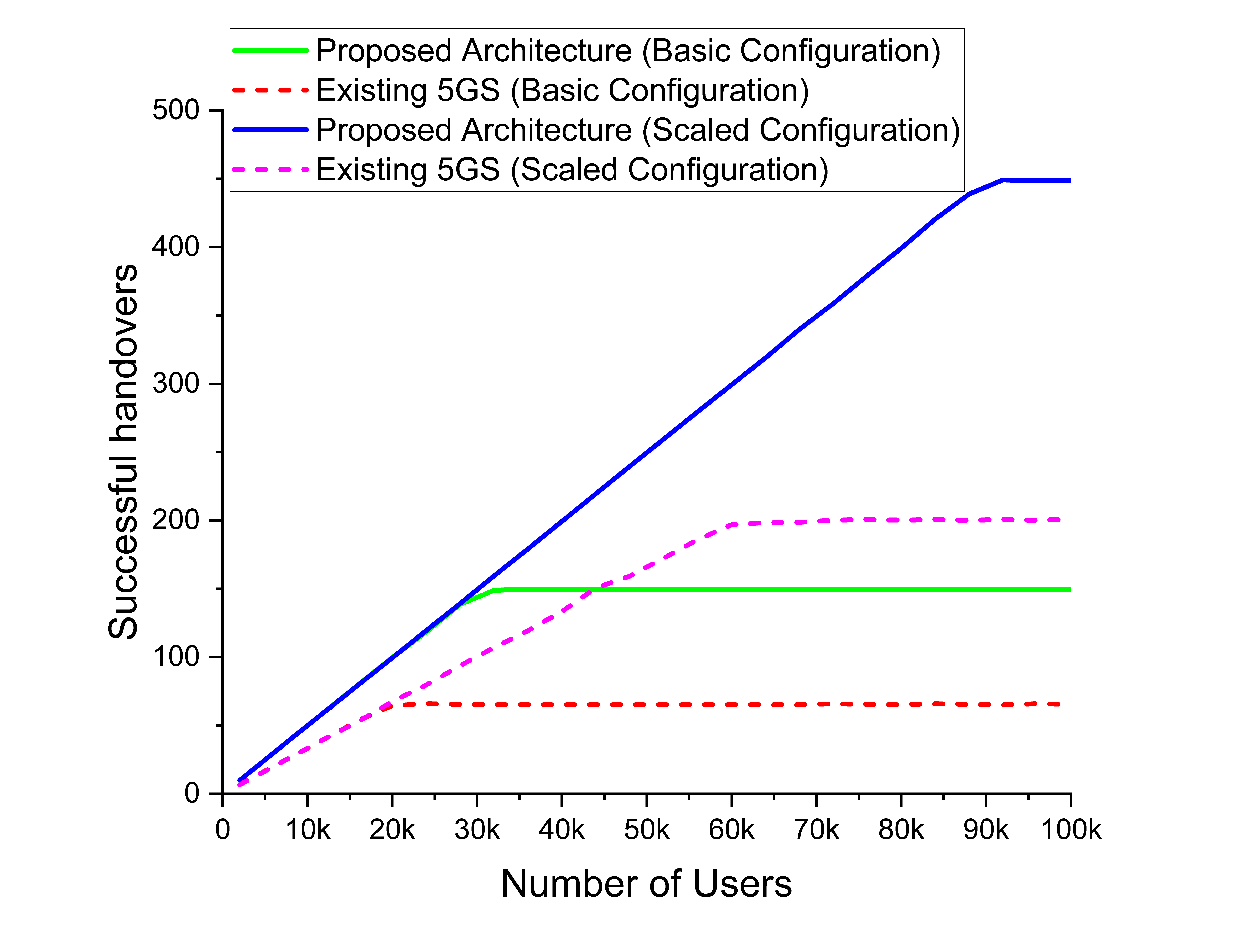}
	\vspace{-0.4cm}
    \caption{Number of successful handovers per unit time for the proposed and the 5GS architecture.}
    \label{fig:fig12}
	\vspace{-0.5cm}
\end{figure}


Scalability for the 5GS and the proposed architecture is evaluated as per Equation \ref{eq:eq1}. It is plotted in Figure \ref{fig:fig11} on the basis of the results obtained for session establishment rate, average response time and utilisation from the PEPA-based simulation. As stated earlier, we consider the two configurations \textit{$m_1$} (basic configuration) and \textit{$m_2$} (scaled configuration) for estimating the scalability metric. Figure \ref{fig:fig11} shows that the 5GS can serve 10,000 users for a basic configuration, and the proposed architecture can serve 20,000 users. Similarly, the 5GS reaches its saturation point at 34,000 users, and the proposed architecture saturates at 62,000 users for scaled configuration. As a result, the curve emphasizes that the proposed architecture has the capacity to support a larger number of users, reaching a saturation point later than that of the 5GS. Besides, the proposed architecture is more scalable with increased users for the same number of NFs/processors. Please note that a similar explanation for all the performance measures (successful handovers, processor utilisation and scalability) holds in the case of user mobility service.

\subsubsection{User Mobility Service}
This section presents the comparative analysis of the 5GS and the proposed architecture for the user mobility service. Similar to the session establishment, the analysis is performed for the basic and the scaled configurations. The basic configuration for the proposed architecture is given as \textit{($N_{upt}$, $N_{msf}$, $N_{ran}$, $N_{cn}$, $N_{upf}$) = (1,2,2,1,1)} and for the 5GS architecture is \textit{($N_{sdu}$, $N_{scu}$, $N_{tdu}$, $N_{tcu}$, $N_{amf}$, $N_{smf}$, $N_{upf}$) } = \textit{(1,1,1,1,1,1,1)}. Similarly, the scaled configuration for the proposed architecture is \textit{($N_{upt}$, $N_{msf}$, $N_{ran}$, $N_{cn}$, $N_{upf}$) = (3,6,6,3,3)} and for the 5GS architecture is given as \textit{($N_{sdu}$, $N_{scu}$, $N_{tdu}$, $N_{tcu}$, $N_{amf}$, $N_{smf}$, $N_{upf}$) } = \textit{(3,3,3,3,3,3,3)}. Here $N_{upt}$, $N_{msf}$, $N_{ran}$, $N_{cn}$, $N_{upf}$ are the number of Target-User Plane (T-UP), MSF, RAN controller, CN controller and UPF respectively in the system model. Similarly, $N_{sdu}$, $N_{scu}$, $N_{tdu}$, $N_{tcu}$, $N_{amf}$, $N_{smf}$, $N_{upf}$ are the number of S-DU, S-CU, T-DU, T-CU, AMF, SMF, and UPF respectively. Please note that for brevity, we have not split S-CU into S-CU-CP and S-CU-UP and T-CU into T-CU-CP and T-CU-UP while modelling the user mobility call flow procedure for the 5GS. Further, we use an equal number of functions and associated processors to the 5GS and the proposed architecture for justified comparison.


\begin{figure}[ht]
	\centering
	\includegraphics[width=0.8\columnwidth]{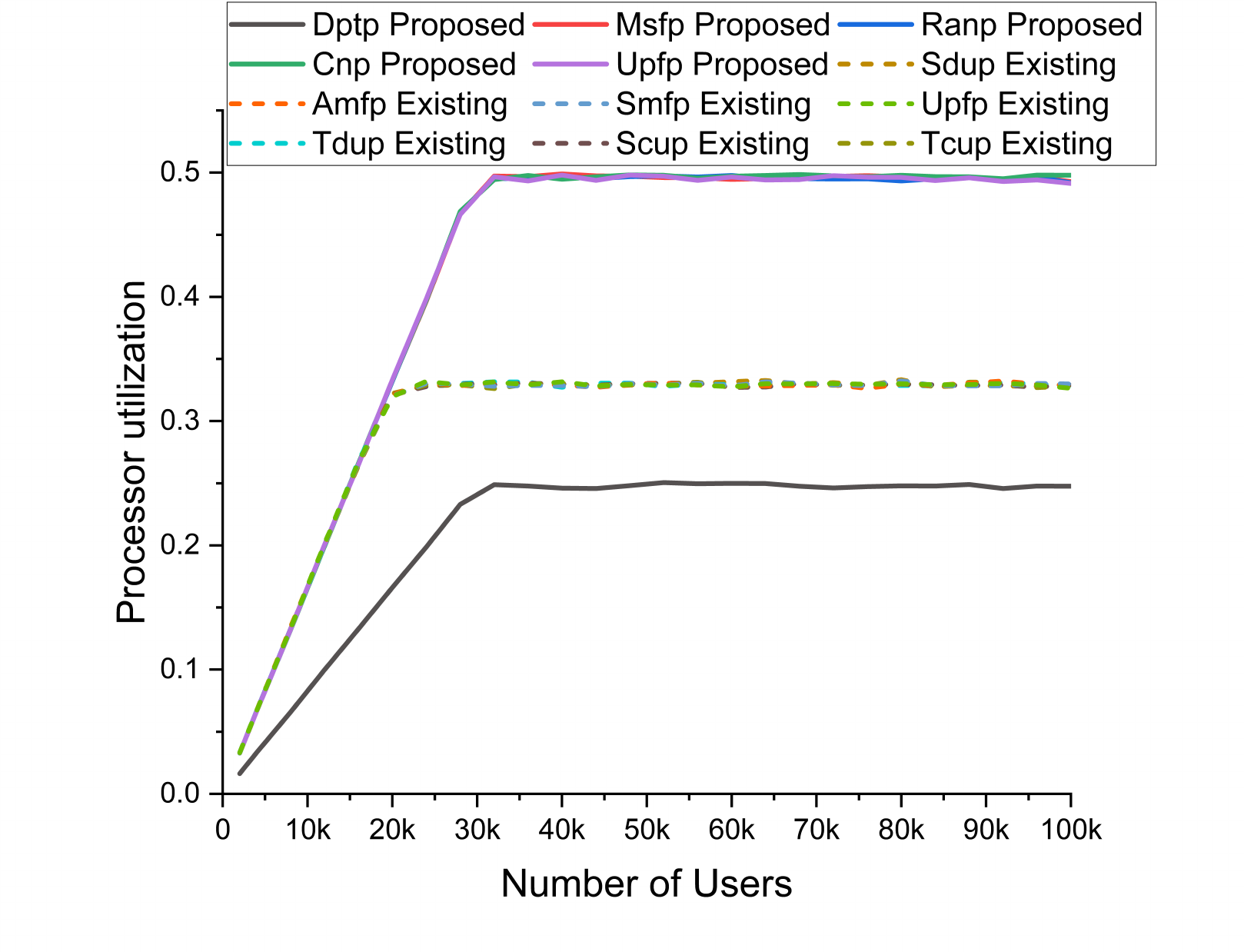}
	\vspace{-0.4cm}
    \caption{Processor utilisation in case of user mobility for the proposed and the 5GS architecture with the basic configuration.}
    \label{fig:fig14}
	\vspace{-0.3cm}
\end{figure}

\begin{figure}[ht]
	\centering
	\includegraphics[width=0.8\columnwidth]{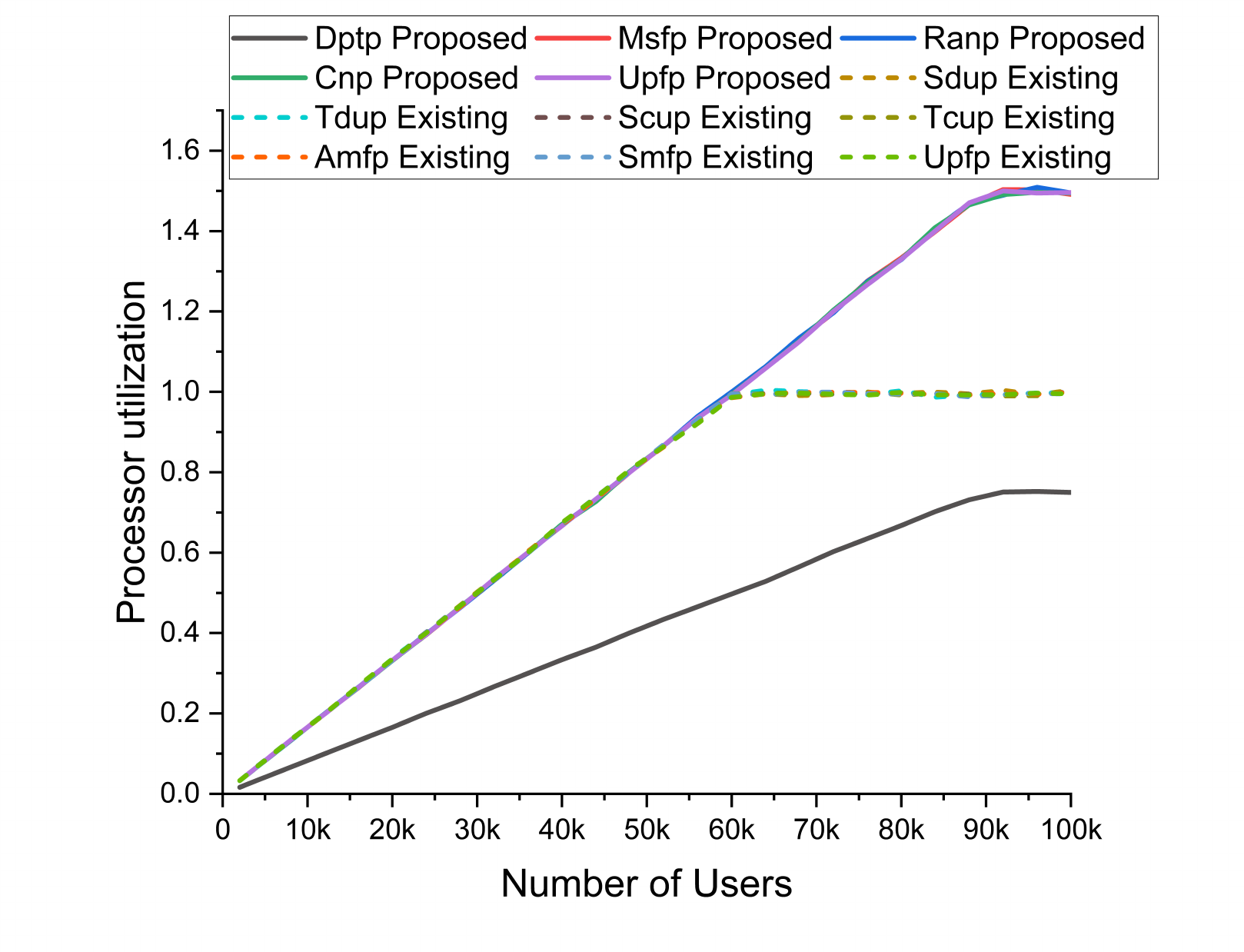}
	\vspace{-0.4cm}
    \caption{Processor utilisation in case of user mobility for the proposed and the 5GS architecture with the scaled configuration.}
    \label{fig:fig15}
	\vspace{-0.2cm}
\end{figure}

\begin{figure}[ht]
	\centering
	\includegraphics[width=0.8\columnwidth]{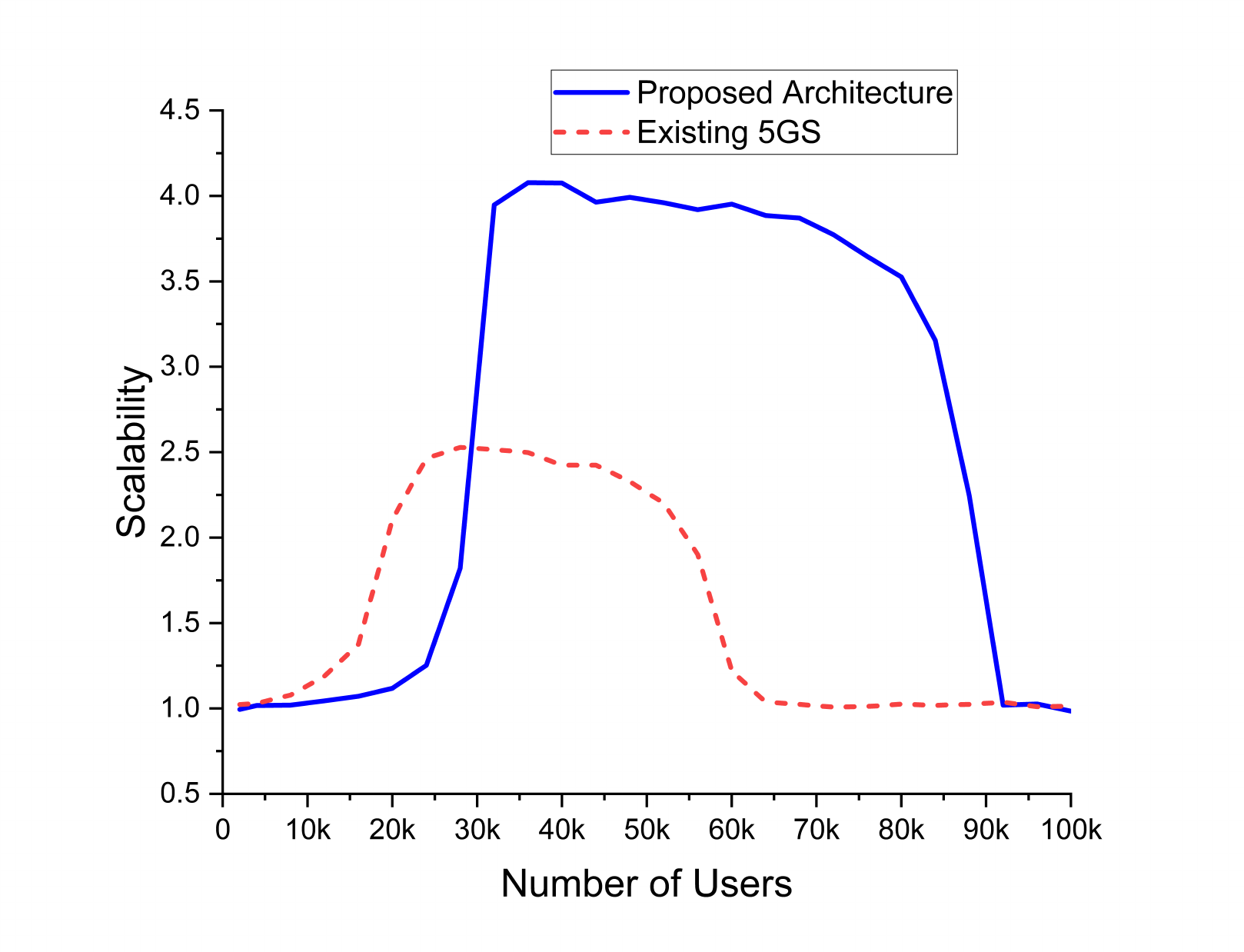}
	\vspace{-0.4cm}
    \caption{Scalability in case of user mobility for the proposed and the 5GS architecture.}
    \label{fig:fig16}
	\vspace{-0.5cm}
\end{figure}

Figure \ref{fig:fig12} shows that the proposed architecture serves more successful handovers per unit time compared to the 5GS for both the basic and the scaled configurations. The saturation point for the 5GS is 20,000 users, while for the proposed architecture, the saturation point is 30,000 users for the basic configuration. Similarly, in the scaled configuration, the saturation point for the 5GS is around 60,000 users, while for the proposed, the saturation is around 90,000 users. The number of successful handovers per unit of time has increased using a scaled configuration for both architectures. 

Figure \ref{fig:fig14} and Figure \ref{fig:fig15} are the results of processor utilisation for both the 5GS and the proposed architecture. A similar observation is noted here as well, indicating that the saturation point for processor utilisation is significantly higher for the proposed architecture viz-a-viz the 5GS. As an illustration, the DPTP reaches its maximum utilisation, elucidating the saturation point for the number of successful handovers per unit time. At this point, other processors are not fully utilised. These findings draw a similar conclusion that the request processing chain fails if an NF becomes a bottleneck in the consecutive chain. Figure \ref{fig:fig16} shows the scalability results in the case of user mobility service for 5GS and the proposed architectures. It can be observed from the scalability results that 5GS reaches its saturation point earlier than the proposed architecture and the proposed architecture is more scalable.

\section{CONCLUSION AND FUTURE WORK}
\label{conc}
In this paper, we have proposed a novel mobile network architecture to separate the handling of UE signalling from the user plane control (resource control) functionality, enhancing the modularity, scalability, and flexibility of the network control plane. The transposition of UE signalling handling from the control plane to service plane is a paradigm shift. It leads to simplified protocols and opens up new ways to implement use case specific signalling in mobile networks. The proposed architecture also has improved implementation of CUPS viz-a-viz 5GS. We have considered PDU session establishment and user mobility services as examples to analyse the performance of the proposed architecture using the PEPA-based simulation method. Based on the performance results and other benefits, it can be concluded that the proposed architecture is a promising option for future networks to handle vast and diverse traffic demands and massive connectivity. In future, we plan to analyse this architecture viz-a-viz its impact on reducing the security threats to the 6G system.


\bibstyle{IEEEtran}
\bibliography{tnsm}

\end{document}